\documentclass{elsart}
\usepackage{amssymb}
\usepackage[german,english]{babel}
\usepackage[intlimits,sumlimits,namelimits]{amsmath}
\usepackage[dvips]{epsfig}
\usepackage[dvips]{color}
\usepackage{ifthen,shadow}
\usepackage{fancyhdr}

\newcommand{\beq}{\begin{equation}}
\newcommand{\eeq}{\end{equation}}
\newcommand{\beqa}{\begin{eqnarray}}
\newcommand{\eeqa}{\end{eqnarray}}
\newcommand{\bra}[1]{\mbox{$\langle #1|$}}
\newcommand{\ket}[1]{\mbox{$|#1\rangle$}}

\begin{document}

\begin{frontmatter}

\title{\bf Polarization observables in the $\gamma d\rightarrow\pi NN$
    reaction in the $\Delta$(1232)-resonance region}

\author{E.M.\ Darwish and Kh.\ Gad}

\address{Physics Department, Faculty of Science, South Valley University, 
  Sohag, Egypt}

\date{\today}

\begin{abstract}
Incoherent pion photoproduction on the deuteron in the 
$\Delta$(1232)-resonance region is investigated with special emphasis 
on polarization observables. For the elementary pion photoproduction operator 
an effective Lagrangian model which includes the standard pseudovector Born 
terms and a resonance contribution 
from the $\Delta$(1232)-excitation is used. Our results for the elementary 
$\gamma N\to\pi N$ reaction are in good agreement with recent experimental 
data and results of other theoretical calculations. A general analysis of 
all possible polarization observables for the $\gamma d \to\pi NN$ reaction 
with polarized photon beam and/or oriented deuteron target is presented. 
The unpolarized differential cross section, photon asymmetry, vector and 
tensor target asymmetries are predicted for forthcoming experiments.

\vspace{0.2cm}

\noindent{\it PACS:}
24.70.+s; 13.60.Le; 25.20.Lj; 21.45.+v.\\
\noindent{\it Keywords:}  Polarization phenomena in reactions; Meson 
production; Photoproduction reactions; Few-body systems.
\end{abstract}
\end{frontmatter}

\section{Introduction} 
\label{sec1}
During the last years, pseudoscalar meson production in
electromagnetic reactions on light nuclei has become a very active
field of research in medium-energy nuclear physics with respect to the
study of hadron structure.  For the following reasons the deuteron
plays an outstanding role besides the free nucleon. The first one is
that the deuteron is the simplest nucleus on whose structure we have
abundant information and a reliable theoretical understanding, i.e.,
the structure of the deuteron is very well understood in comparison to
heavier nuclei. Furthermore, the small binding energy of nucleons in
the deuteron, which from the kinematical point of view provides the
case of a nearly free neutron target, allows one to compare the
contributions of its constituents to the electromagnetic and hadronic
reactions to those from free nucleons in order to estimate interaction
effects.

Meson photo- and electroproduction on light nuclei is primarily
motivated by the following possibilities: (i) study of the elementary
neutron amplitude in the absence of a neutron target, (ii)
investigation of medium effects, i.e., possible changes of the
production operator in the presence of other nucleons, (iii) it
provides an interesting means to study nuclear structure, and (iv) 
it gives information on pion production on off-shell nucleon, as 
well as on the very important $\Delta N$-interaction in a nuclear 
medium. As an
illustration of these various aspects, we will investigate in this
paper incoherent pion photoproduction on the deuteron in the
$\Delta$(1232)-resonance region with special emphasis on polarization
observables. The importance of this process derives from the fact that
the deuteron, being the simplest nuclear system, plays a similar
fundamental role in nuclear physics as the hydrogen atom plays in
atomic physics.

The major reason for studying polarization phenomena in reactions of
the type $a+b\to c+d+...$ lies in the fact that only the use of
polarization degrees of freedom allows one to obtain complete
information on all possible reaction matrix elements. Without
polarization, the cross section is given by the incoherent sum of
squares of the reaction matrix elements only. Thus, small amplitudes
are masked by the dominant ones. On the other hand, small amplitudes
very often contain interesting information on subtle dynamical
effects. This is the place where polarization observables enter,
because such observables in general contain interference terms of the
various matrix elements in different ways. Thus, a small amplitude may
be considerably amplified by the interference with a dominant matrix
element. An example is provided by the influence of the small electric
form factor of the neutron on the transverse in-plane component of the
neutron polarization in quasifree deuteron
electrodisintegration~\cite{Arn81,ArenZ88,Klein96}. It is just this
feature for which polarization physics has become such an important
topic in various branches of physics.

The present paper
is a natural extension of our work in~\cite{Dar03,Dar03+} 
where we have presented the energy dependence of the $\gamma
d\rightarrow\pi NN$ reaction over the whole $\Delta$(1232)-resonance region and
gave results for differential and total cross sections as well as
results for the spin asymmetry and Gerasimov-Drell-Hearn (GDH) sum
rule for the deuteron.  Notwithstanding this continuing effort to
study this process, the wealth of information contained in it has not
yet been fully exploited. Since the $t$-matrix has 12 independent
complex amplitudes, one has to measure 23 independent observables, in
principle, in order to determine completely the $t$-matrix. Up to
present times, only a few observables have been measured and studied in
detail, e.g., differential and total cross sections.

On the other hand, the measurement of polarization observables is a
rather difficult task, requiring great experimental skill and advanced
technology, which explains why little is known about these more
involved polarization observables. However, in view of the recent
technical improvements, e.g., at MAMI in Mainz, ELSA in Bonn and JLab 
in Newport News, for
preparing polarized beams and targets and for polarimeters for the
polarization analysis of ejected particles it appears timely to study
in detail polarization observables in pion production on the deuteron.
The aim will be to see what kind of information is buried in the
various polarization observables, in particular, what can be learned
about the role of subnuclear degrees of freedom like meson and isobar
or even quark-gluon degrees of freedom.

The paper is organized as follows. In Section \ref{sec2} we will
present the effective Lagrangian model for the elementary pion
photoproduction amplitude which will serve as an input for the
reaction on the deuteron. Its predictions for differential and total
cross sections are compared with recent experimental data and results 
of other theoretical predictions. Section
\ref{sec3} will introduce the general form of the differential cross
section for incoherent pion photoproduction on the deuteron. The
treatment of the $\gamma d\to\pi NN$ amplitude, based on time-ordered
perturbation theory, will be described in Section \ref{sec4}.  In Section
\ref{sec5} we will give the complete formal expressions of
polarization observables for the $\gamma d\to\pi NN$ reaction with polarized 
photon beam and/or oriented deuteron target in terms
of the $t$-matrix elements. Details of the actual calculation and the
results will be presented and discussed in Section \ref{sec6}. Finally, we
close in Section \ref{sec7} with a conclusion.

\section{The $\gamma N\to\pi N$ amplitude}
\label{sec2}
\begin{figure}[htp]
\begin{center}
\includegraphics[scale=0.3]{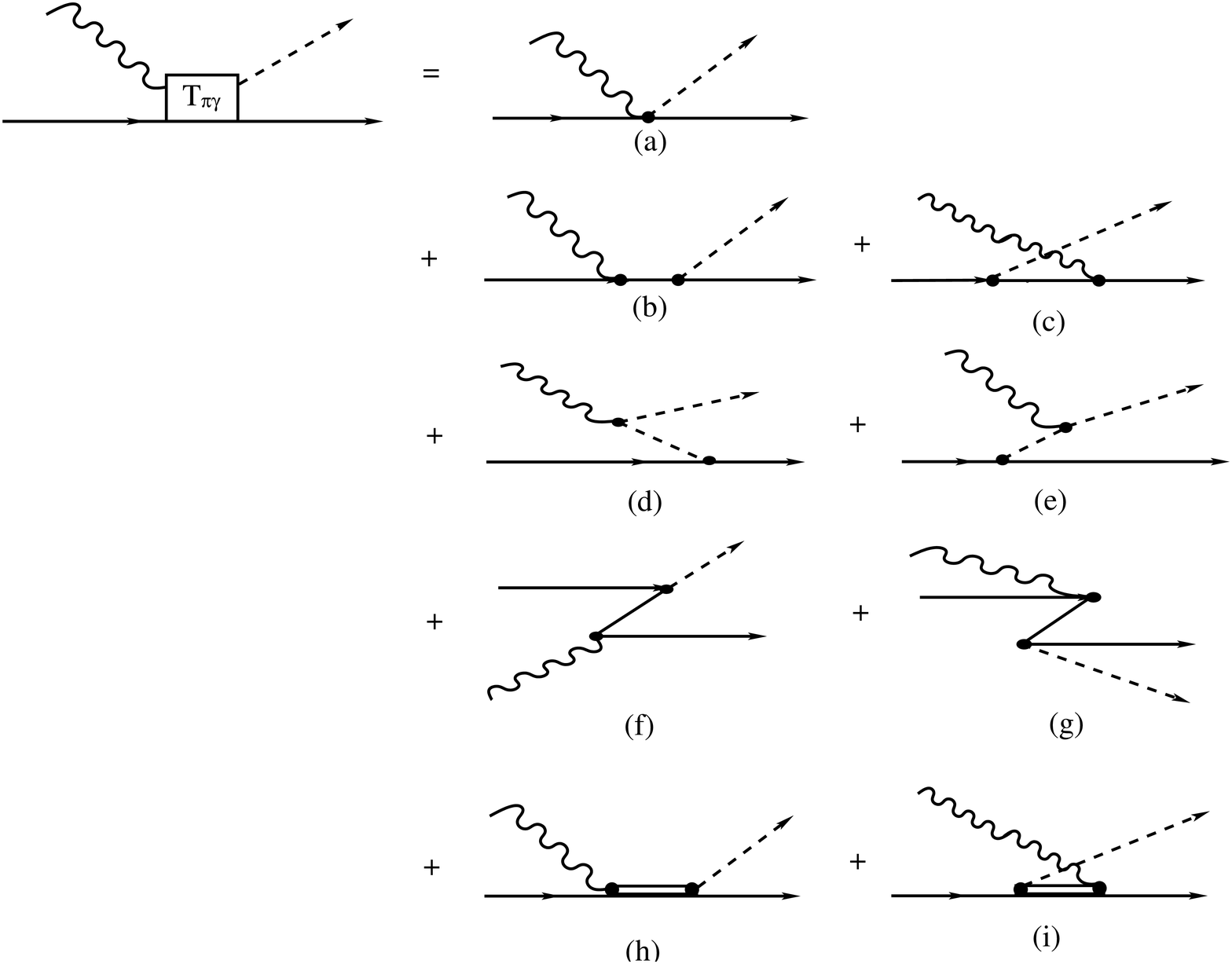}
\caption{Diagrams for the elementary process $\gamma N
    \rightarrow \pi N$: (a) the Kroll-Rudermann graph, (b) and
    (c) the two time-ordered contributions to the direct and crossed
    nucleon pole graph, (d) and (e) the two time-ordered contributions
    to the pion pole graph, (f) and (g) the Z-graphs, and (h) and (i) the
    $\Delta(1232)$-resonance graphs. A solid, dashed and wavy lines 
    represent a nucleon, pion and photon, respectively.}
\label{Fdiagrams}
\end{center}
\end{figure}

\begin{figure}[htb]
\begin{center}
\includegraphics[scale=.7]{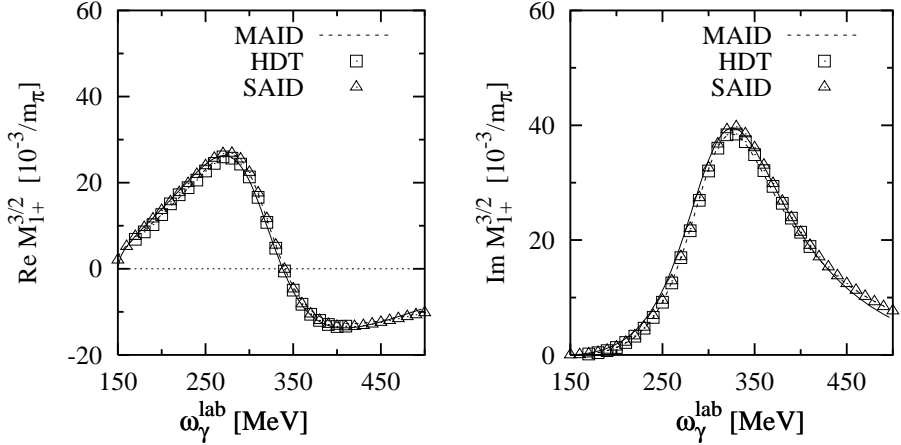}
\vspace*{-0.4cm}
\caption{Real and imaginary parts of the $M_{1+}^{3/2}$
multipole. Notation: solid curves: present model; dotted curves: 
MAID~\protect\cite{Maid}. Data points: from~\protect\cite{Said}
(SAID, solution: September 2000), \protect\cite{HaD98} (HDT).}
\label{multipoles}
\end{center}
\end{figure}

For the elementary pion photoproduction operator, we have taken, as in
our previous work~\cite{Dar03,Dar03+}, the effective
Lagrangian model of Schmidt {\it et al.}~\cite{ScA96}. This model had
been constructed to give a realistic description of the 
$\Delta$(1232)-resonance region.  It is given in an arbitrary frame of reference and
allows a well defined off-shell continuation as required for studying
pion production on nuclei.  It is in contrast to other approaches,
where the elementary amplitude is constructed first on-shell in the
photon-nucleon center-of-mass (c.m.) frame with subsequent boost into an 
arbitrary
reference frame and some prescription for the off-shell continuation.
In the latter method, one loses terms which by chance vanish in the
c.m.\ frame~\cite{BrA97}. In our approach, the only uncertainty arises
from the assignment of the invariant energy for the photon-nucleon
subsystem in the resonance propagators as has been discussed in detail
in~\cite{BrA97}. Here we use the spectator on-shell approach. The
model of Schmidt {\it et al.}~\cite{ScA96} consists of the standard 
pseudovector Born terms
and the contribution of the $\Delta(1232)$-resonance. The individual
terms of the matrix element for pion photoproduction reaction on the
free nucleon are shown in Fig.~\ref{Fdiagrams}.  For further details
with respect to the elementary pion photoproduction operator we refer
to~\cite{ScA96}. The parameters of the $\Delta$-resonance are fixed
by fitting the experimental $M_{1+}^{3/2}$ multipole which is dominant 
in the region of the $\Delta$-resonance. The quality of
the model can be judged by a comparison with the MAID 
analysis~\cite{Maid}, the Mainz dispersion analysis~\cite{HaD98}, and
the VPI analysis~\cite{Said} as shown in Fig.~\ref{multipoles}, and
one notes quite a good agreement.

\begin{figure}[htb]
\begin{center}
\includegraphics[scale=.7]{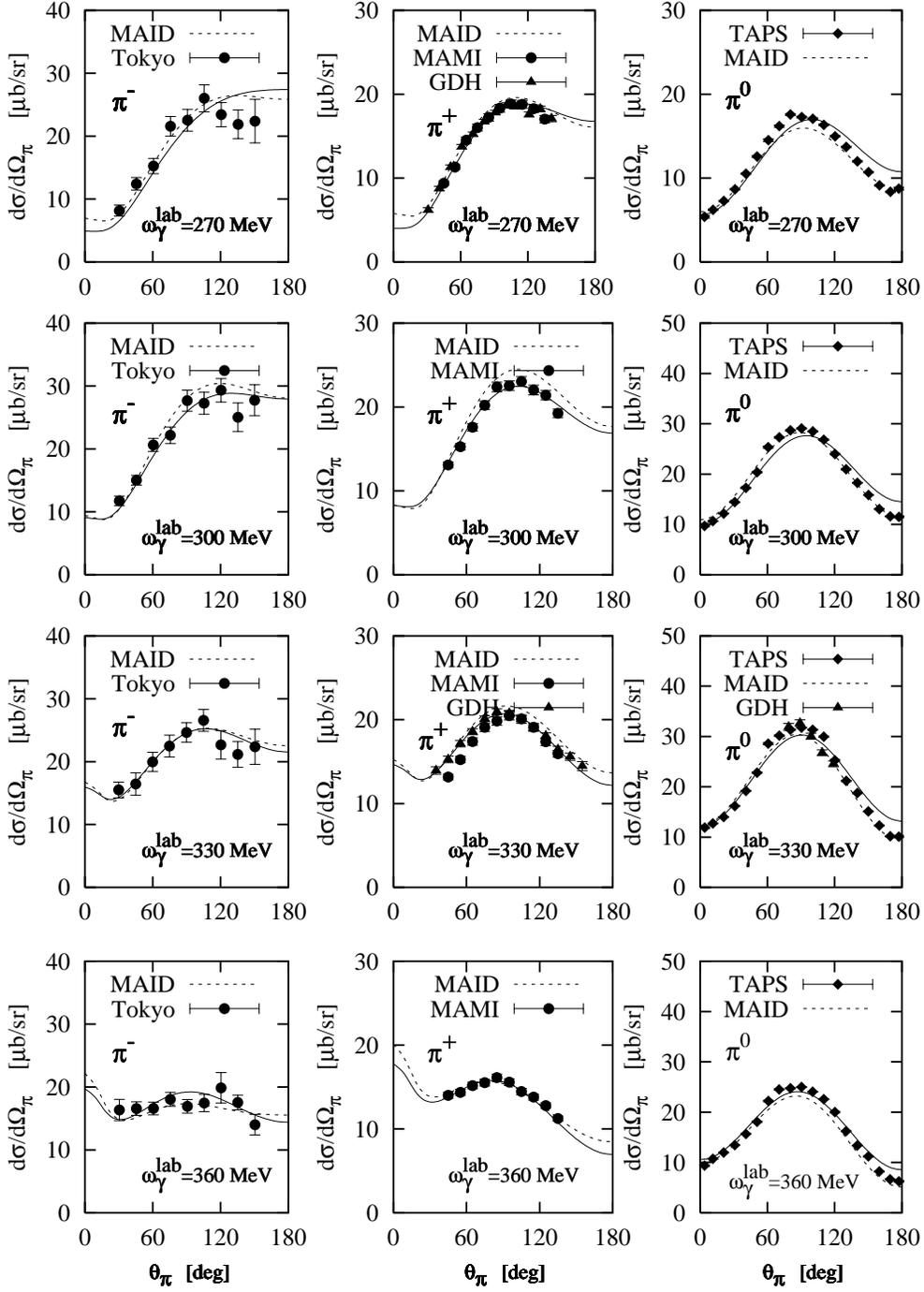}
\caption{Differential cross section for the elementary reaction on 
the nucleon for the three charge states of the pion at various  
photon energies. Left panels: $\pi^-$, middle panels: $\pi^+$, and 
right panels: $\pi^0$. The solid curves: present model; dotted curves: 
MAID~\protect\cite{Maid}. Experimental data from
\protect\cite{Fu+77} (Tokyo) for $\pi^-$, \protect\cite{Leu00} (MAMI),
\protect\cite{Pre00} (GDH) for $\pi^0$, and \protect\cite{Be+97} (TAPS),
\protect\cite{Pre00} (GDH) for $\pi^+$.}  
\label{diff_elem}
\end{center}
\end{figure}

In Fig.~\ref{diff_elem} we compare our results for the differential
cross sections with the MAID analysis~\cite{Maid} and with
experimental data. For $\pi^{+}$ and $\pi^{0}$ photoproduction on the
proton the data are taken from~\cite{Be+97} (TAPS), \cite{Leu00} (MAMI) 
and \cite{Pre00} (GDH),
whereas for $\pi^{-}$ photoproduction on the neutron we took the data
of the inverse reaction $\gamma n\leftarrow p\pi^-$ from~\cite{Fu+77}
(Tokyo).  In general, we obtain quite a good agreement with the data,
especially in the region of the $\Delta$(1232)-resonance (330 MeV).
Also in comparison with the MAID analysis our elementary production
operator does quite well in this energy region. One notes only small
discrepancies which very likely come from the fact that no other
resonances besides the $\Delta$(1232) are included in our model.

\begin{figure}[htb]
\begin{center}
\includegraphics[scale=.7]{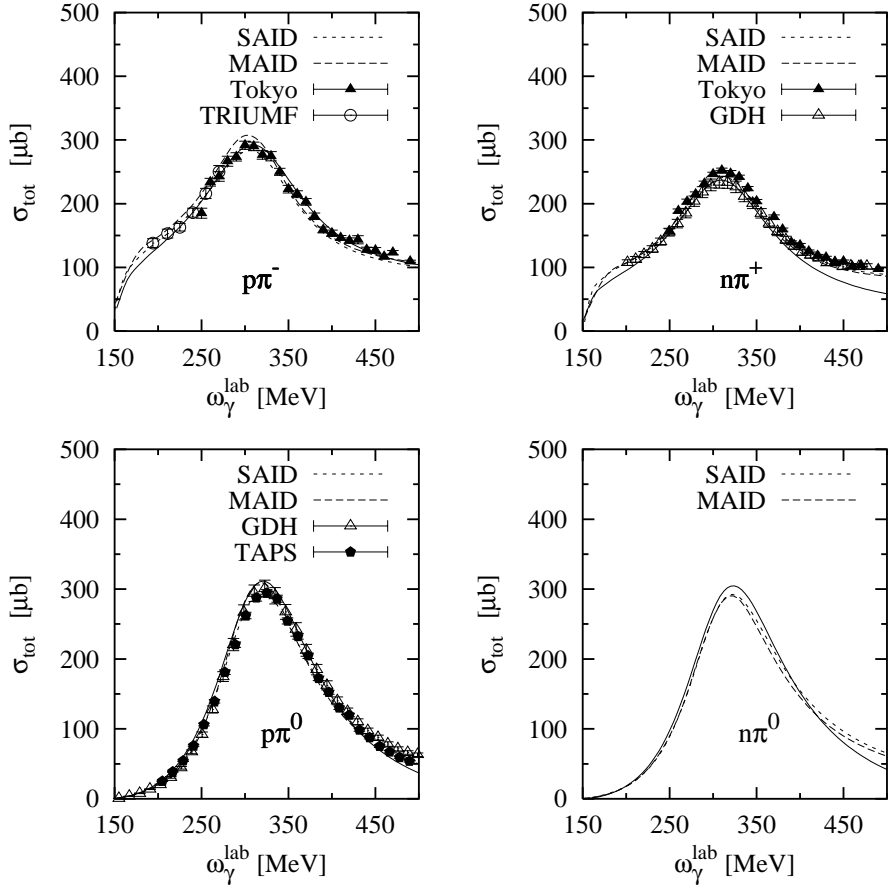}
\vspace*{-0.4cm}
\caption{Total cross sections for pion photoproduction on the nucleon
as a function of photon energy for all four physical
channels. Notation of the curves: solid: present model; dotted:
SAID~\protect\cite{Said}; dashed: MAID~\protect\cite{Maid}. Experimental 
data from \protect\cite{Fu+77} (Tokyo), \protect\cite{Bagheri88} (TRIUMF),
\protect\cite{Pre00} (GDH), \protect\cite{Hae96} (TAPS).} 
\label{tot}
\end{center}
\end{figure}

The total cross sections for the different pion channels are shown in
Fig.~\ref{tot} and compared with experimental data. In general, we
obtain a good agreement with the data using the small value ${f^2_{\pi
    N}}/{4\pi}=0.069$ for the pion-nucleon coupling constant. The
agreement with the data from~\cite{Fu+77} and~\cite{Bagheri88} for
$\pi^-$ photoproduction is again satisfactory. In case of the $\pi^+$
photoproduction, the agreement is good up to a photon energy of about
400 MeV. For higher energies, the $D_{13}$-resonance, which is not
included in our calculation, gives a non-vanishing
contribution~\cite{Maid}. The $\pi^{+}$ data from~\cite{Fu+77} are
slightly underestimated in the resonance region by our calculation but
also by the MAID analysis. Except for a tiny overestimation in the
maximum, the description of the data from~\cite{Pre00,Hae96} for $\pi^0$ 
production on the proton is also very
good. Therefore, this model for the elementary photoproduction
amplitude is quite satisfactory for our purpose, namely to incorporate
it into the reaction on the deuteron.

\section{Process on the deuteron}
\label{sec3}
In this section we will briefly review the general formalism for
incoherent pion photoproduction on the deuteron. The general
expression for the 5-fold unpolarized differential cross section of pion
photoproduction reaction on the deuteron is given, using the
conventions of Bjorken and Drell~\cite{BjD64}, by
\beqa
d\sigma = (2\pi)^{-5}\delta^{4}\left( k+d-p_{1}-p_{2}-q\right)
\frac{1}{|\vec{v}_{\gamma}-\vec{v}_{d}|} \frac{1}{2}
\frac{d^{3}q}{2\omega_{\vec{q}}} \frac{d^{3}p_{1}}{E_{1}}
\frac{d^{3}p_{2}}{E_{2}}
\frac{M_{N}^{2}}{4\omega_{\gamma}E_{d}}
\nonumber \\ & & \hspace{-9cm}\times~
\frac{1}{6}\sum_{\alpha} 
|{\mathcal M}^{(t\mu)}_{s m m_{\gamma} m_d}|^{2} \, , 
\eeqa
where we have introduced as a shorthand for the quantum numbers 
$\alpha=(s,m,t,m_{\gamma},m_d)$. The initial photon and deuteron four-momenta are denoted by
$k=(\omega_\gamma,\vec k\,)$ and $d=(E_d,\vec d\,)$, respectively, and
the four-momenta of final meson and two nucleons by $q=(\omega_q,\vec
q\,)$ with $\omega_{q} = \sqrt{m_{\pi}^{2} + \vec{q}^{\,2}}$, $m_\pi$
as pion mass, and $p_j=(E_j,\vec p_j\,)$ ($j=1,2$) with $E_{j}
=\sqrt{M_{N}^{2} + \vec{p}_{j}^{\,2}}$, respectively, and $M_N$ as
nucleon mass. Furthermore, $m_{\gamma}$ denotes the photon
polarization, $m_{d}$ the spin projection of the deuteron, $s$ and $m$
total spin and projection of the two outgoing nucleons, respectively,
$t$ their total isospin, $\mu$ the isospin projection of the pion, and
$\vec{v}_{\gamma}$ and $\vec{v}_{d}$ the velocities of photon and
deuteron, respectively.  The states of all particles are covariantly
normalized.  The reaction amplitude is denoted by ${\mathcal
  M}^{(t\mu)}_{sm m_{\gamma}m_d}$. As
in~\cite{Dar03,Dar03+,ScA96}, we have chosen as independent
variables the pion momentum $q$, its angles $\theta_{\pi}$ and
$\phi_{\pi}$, the polar angle $\theta_{p_{NN}}$ and the azimuthal
angle $\phi_{p_{NN}}$ of the relative momentum $\vec p_{NN}$ of the
two outgoing nucleons as independent variables. We prefer this choice
of variables, because in this case the kinematic factor do not has any
singularity on the boundary of the available phase space, when $p_{NN}\to
0$.

The total and relative momenta of the final $NN$-system are defined  
respectively by
\beqa
\vec{P}_{NN} = \vec{p}_{1} + \vec{p}_{2}= \vec{k} - \vec{q}\,
\eeqa
and
\beqa
\vec p_{NN} = \frac{1}{2}\left(\vec{p}_{1} - \vec{p}_{2}\right)\, .
\eeqa
The absolute value of the relative momentum $\vec p_{NN}$ is given by 
\beqa
\label{relm}
p_{NN} = \frac{1}{2}\sqrt{ \frac{ E_{NN}^{2}(W_{NN}^{2}-4
M_{N}^{2}) }{E_{NN}^{2}-P^{2}_{NN}\cos^{2}\theta_{Pp_{NN}}} }\, ,
\eeqa
where $\theta_{Pp_{NN}}$ is the angle between $\vec{P}_{NN}$ and 
$\vec p_{NN}$. $E_{NN}$ and $W_{NN}$ denote total energy and invariant mass of 
the $NN$-subsystem, respectively, and are given by
\beqa
E_{NN} & = & E_{1}+E_{_2}=  \omega_{\gamma}+E_{d}-\omega_{q} \, , \nonumber \\
W_{NN}^{2} & = & {E}_{NN}^{2}-P^{2}_{NN}\,.
\eeqa

\begin{figure}[htb]
\begin{center}
\includegraphics[scale=1.0]{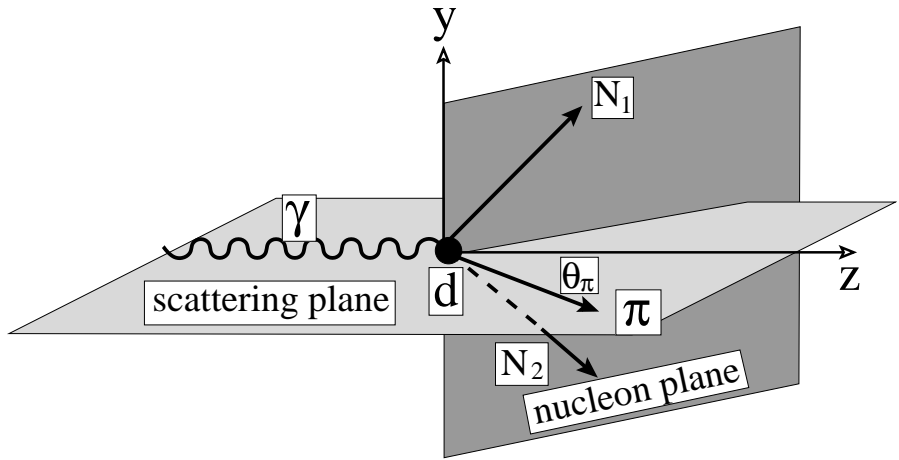}
\caption{\small Kinematics in the laboratory system for $\gamma d\rightarrow 
  \pi NN$.}
\label{labsys}
\end{center}
\end{figure}

For the evaluation we have chosen the laboratory frame where
$d^{\mu}=(M_d,\vec 0\,)$ with $M_d$ as deuteron mass. As coordinate
system a right-handed one is taken with $z$-axis along the
momentum $\vec k$ of the incoming photon and $y$-axis along 
$\vec k\times\vec q$. Thus, the outgoing pion defines the
scattering plane. Another plane is defined by the momenta of the 
outgoing nucleons which we will call the nucleon plane 
(see Fig.~\ref{labsys}). 

The fully exclusive differential cross section is given by
\beqa
\frac{d^5\sigma}{d\Omega_{p_{NN}} d\Omega_{\pi} dq} = \frac{\rho_{s}}{6}
\sum_{\alpha} |{\mathcal M}^{(t\mu)}_{sm m_{\gamma}m_d}|^{2}\,,
\label{fivefold}
\eeqa
where the phase space factor $\rho_{s}$ is expressed in terms of relative and 
total momenta of the two final nucleons 
\beqa
\label{rhos}
\rho_{s} & = & \frac{1}{(2\pi)^{5}}\frac{p_{NN}^{2}M_{N}^{2}}
{\left| E_{2} (p_{NN}+\frac{1}{2} P_{NN} 
\cos\theta_{Pp_{NN}}) + E_{1}
(p_{NN}-\frac{1}{2} P_{NN} \cos\theta_{Pp_{NN}}) \right| } \nonumber \\
& & \times~\frac{q^{2}}{16\omega_{\gamma}M_{d}\omega_{q}} \, .
\eeqa

\section{The $\gamma d\to\pi NN$ amplitude}
\label{sec4}
A plane wave impulse approximation usually serves as the starting
point to calculate the amplitude for electromagnetic pion production
on the deuteron \cite{Osl76,Laz76,Bos78,Rek91} or on a nucleus in
general. It corresponds to a direct embedding of the elementary
amplitudes into the two-nucleon system. The general form of the
photoproduction transition matrix is given by
\beqa
{\mathcal M}^{(t\mu)}_{sm m_{\gamma}m_d}(\vec{k},\vec{q},\vec{p_1},\vec{p_2})
& = & ^{(-)}\bra{\vec{q}\,\mu,\vec{p_1}\vec{p_2}\,s\,m\,t-\mu}\epsilon_{\mu}
(m_{\gamma})J^{\mu}(0)\ket{\vec{d}\,m_d\,00}\, , 
\eeqa
where $J^{\mu}(0)$ denotes the current operator and
$\epsilon_{\mu}(m_{\gamma})$ the photon polarization vector. The
electromagnetic interaction consists of the elementary production
process on one of the nucleons $T_{\pi\gamma}^{(j)}$ $(j=1,2)$ and in
principle a possible irreducible two-body production operator
$T_{\pi\gamma}^{(NN)}$. The final $\pi NN$ state is then subject to
the various hadronic two-body interactions as described by an
half-off-shell three-body scattering amplitude $T^{\pi NN}$. In the
following, we will neglect the electromagnetic two-body production
$T_{\pi\gamma}^{(NN)}$ and the outgoing $\pi NN$ scattering state is
approximated by the free $\pi NN$ plane wave, i.e.,
\beqa
\ket{\vec{q}\,\mu,\vec{p_1}\vec{p_2}\,s\,m\,t-\mu}^{(-)}=
\ket{\vec{q}\,\mu,\vec{p_1}\vec{p_2}\,s\,m\,t-\mu}\,.
\eeqa
This means, we include only the pure plane wave impulse approximation (IA), 
which is defined by the electromagnetic pion production on one of the
nucleons alone, while a more realistic treatment including final state
interaction as well as two-body effects will be reported in a
forthcoming paper. Our justification for such a procedure is the fact
that the IA is the primary process which will be gauged against all other 
effects.

For the spin $( |s m\rangle )$ and isospin $( |t -\mu\rangle )$ part
of the two nucleon wave functions we use a coupled spin-isospin basis
$\ket{s m,\,t -\mu}$.  The antisymmetric final $NN$ plane wave function
thus has the form
\beq
  |\vec{p}_{1},\vec{p}_{2},s m,t -\mu \rangle =
  \frac{1}{\sqrt{2}}\left(
    |\vec{p}_{1}\rangle^{(1)}|\vec{p}_{2}\rangle^{(2)} - (-)^{s+t}
    |\vec{p}_{2}\rangle^{(1)}|\vec{p}_{1}\rangle^{(2)}\right)|s
  m\,,t -\mu\rangle\,,
\eeq
where the superscript indicates to which particle the ket refers.
In the case of charged pions, only the $t = 1$ channel contributes 
whereas for $\pi^{0}$ production both $t = 0$ and $t = 1$ channels
have to be taken into account. Then, the matrix element is given by 
\beqa
  {\mathcal M}_{sm m_{\gamma}m_d}^{(t\mu)}
  (\vec{k},\vec{q},\vec{p_1},\vec{p_2})&=& \langle
\vec{p}_{1},\vec{p}_{2},s m,t -\mu 
  |t_{\gamma\pi}^{NN}(\vec k,\vec q\,)|\vec{d}m_{d},00 \rangle \nonumber\\
&=&\frac{1}{2} \int
  \frac{d^{3}p^{\prime}_{1}}{(2\pi)^{3}} \int
  \frac{d^{3}p^{\prime}_{2}}{(2\pi)^{3}}
  \frac{M_{N}^2}{E^{\,\prime}_{1}E^{\,\prime}_2}\nonumber\\
 && \times \sum_{m^{\prime}} \langle \,\vec{p}_{1}\vec{p}_{2},s
  m,t  -\mu |\, t_{\gamma\pi}^{NN}(\vec k,\vec q\,)
  |\,\vec{p}_{1}^{\,\prime}\vec{p}_{2}^{\,\prime}, 1 m^{\prime}, 0
  0 \rangle \nonumber \\
&& \times \langle\,\vec{p}_{1}^{\,\prime}\vec{p}_{2}^{\,\prime}, 1
  m^{\prime},00|\,\vec{d} m_{d},\,00\rangle
\eeqa
with 
\beqa
t_{\gamma\pi}^{NN}(\vec k,\vec q\,)=t_{\gamma\pi}^{N(1)}(\vec k,\vec q\,)
+t_{\gamma\pi}^{N(2)}(\vec k,\vec q\,)\,,\label{tmat-IA}
\eeqa
where $t_{\gamma\pi}^{N(j)}$ denotes the elementary production amplitude 
on nucleon ``$j$''. As mentioned above, we use covariant normalization for
the nucleon, deuteron and meson states, i.e.,
\beqa
\langle \vec{p}^{\,\prime} |\, \vec{p}\,\rangle & = &
(2\pi)^{3}\frac{E_{p}}{M_{N}}
\,\delta^{3}\!\left( \vec{p}^{\,\prime}-\vec{p}\,\right)\,,\quad
\langle \vec{d}^{\,\prime} |\, \vec{d}\,\rangle = (2\pi)^{3} 2E_{d}
\,\delta^{3}\,( \vec{d}^{\,\prime}-\vec{d}\,)\,, \nonumber \\
\langle \vec{q}^{\,\prime} |\, \vec{q}\,\rangle &=& (2\pi)^3\,2\,
\omega_q\,\delta(\vec q^{\,\prime}-\vec q)\,.
\eeqa
The deuteron wave function has the form
\beq
  \langle\,\vec{p}_{1}\vec{p}_{2}, 1
  m,\,00|\,\vec{d} m_{d},00\rangle = (2\pi)^{3}
  \delta^{3}(\,
    \vec{d}-\vec{p}_{1}-\vec{p}_2 \,)
  \frac{\sqrt{2\,E_{1}E_{2}}}
  {M_{N}} \widetilde{\Psi}_{m,m_{d}}(\vec{p}_{NN})
\eeq
with
\beqa
  \widetilde{\Psi}_{m,m_{d}}(\vec{p}\,) =
  (2\pi)^{\frac{3}{2}}\sqrt{2E_{d}}
  \sum_{L=0,2}\sum_{m_{L}}i^{L}\,C^{L 1 1}_{m_{L} m m_{d}}\,
  u_{L}(p)Y_{Lm_{L}}(\hat{p}) \,,
\eeqa
denoting with $C^{j_1 j_2 j}_{m_1 m_2 m}$ a Clebsch-Gordan coefficient, 
$u_{L}(p)$ the radial deuteron wave function and $Y_{Lm_{L}}(\hat{p})$ 
a spherical harmonics. Using (\ref{tmat-IA}) one finds in the laboratory 
system for the matrix element the following expression
\beqa
\label{tmat_IA_lab}
  {\mathcal M}_{sm m_{\gamma}m_d}^{(t\mu)}
  (\vec k,\vec q,\vec p_1,\vec p_2) &=&
 \sqrt{2}\sum_{m^{\prime}}\langle s 
  m,\,t -\mu|\,\Big( \langle
  \vec{p}_{1}|t_{\gamma\pi}^{N(1)}(\vec k,\vec q\,)|-\vec{p}_{2}\rangle
  \tilde{\Psi}_{m^{\prime},m_{d}}(\vec{p}_{2}) \nonumber\\  
\hspace{1cm}& &
-(-)^{s+t}(\vec p_1 \leftrightarrow \vec p_2) 
\Big)\,|1 m^{\prime},\,00\rangle \,.
\eeqa
Note, that in (\ref{tmat_IA_lab}) the elementary production operator acts on nucleon
``1'' only. This matrix element possesses the obvious symmetry under the
interchange of the nucleon momenta
\beqa
  {\mathcal M}_{sm m_{\gamma}m_d}^{(t\mu)}
  (\vec k,\vec q,\vec p_2,\vec p_1) =(-)^{s+t+1}\,
  {\mathcal M}_{sm m_{\gamma}m_d}^{(t\mu)}
  (\vec k,\vec q,\vec p_1,\vec p_2) \,.
\eeqa

Choosing the $z$-axis in the direction of the incoming photon and isolating
the azimuthal dependence of the direction of pion momentum, we obtain the
following general form for the reaction matrix  
\beqa
\mathcal M^{(t\mu)}_{sm,m_{\gamma}m_d}(\vec k,\vec q,\vec p_1,\vec p_2) & = &
\mathcal
T^{(t\mu)}_{sm,m_{\gamma}m_d}(\vec k,q,\theta_{\pi},\vec p_1,\vec p_2)
~e^{i(m_{\gamma}+m_d)\phi_{\pi}}\, . 
\eeqa
Using parity conservation one can shows, that
the reduced $\mathcal T$-matrix elements obey the following symmetry relation
\beqa
\mathcal T^{(t\mu)}_{s,-m,-m_{\gamma},-m_d} & = &
(-)^{s+m+m_{\gamma}+m_d} ~\mathcal T^{(t\mu)}_{s,m,m_{\gamma},m_d}\, .
\eeqa
This symmetry relation reduces the number of complex amplitudes from 24 to 12
independent ones. For their determination one needs 23 real observables since
a overall phase remains arbitrary.

\section{Polarization observables}
\label{sec5}
Polarization observables will give additional valuable information for 
checking the spin degrees of freedom of the elementary pion production 
amplitude of the neutron, provided, and this is very important, that 
one has under control all interfering interaction effects which prevent 
a simple extraction of this amplitude. For the definition of these 
observables in terms of the transition matrix elements we refer the 
reader to~\cite{Aren88}, in which all possible polarization 
observables in $d(\gamma,N)N$ with polarized photon beam and/or oriented 
deuteron target are derived. We briefly recall here the necessary notations 
and definitions. 

The cross section for arbitrary polarized photons and initial deuterons can 
be computed for a given $\mathcal M$-matrix by applying the density matrix 
formalism similar to that given by Arenh\"ovel~\cite{Aren88} for deuteron
photodisintegration. The most general expression for all possible polarization 
observables is given by 
\beqa
\mathcal O &=& {\rm Tr}~(\mathcal M^{\dag}~\Omega~\mathcal M~\rho)\nonumber \\
& = & \sum_{\alpha\alpha^{\prime}} \int d\Omega_{p_{NN}}~\rho_s~
\mathcal M^{(t^{\prime}\mu^{\prime})~\star}_{s^{\prime}m^{\prime},
m_{\gamma}^{\prime}m_d^{\prime}}~\vec{\Omega}_{s^{\prime}m^{\prime}sm}
~\mathcal M^{(t\mu)}_{sm,m_{\gamma}m_d}
~\rho^{\gamma}_{m_{\gamma}m_{\gamma}^{\prime}}~ \rho^{d}_{m_dm_{d}^{\prime}}\,,
\eeqa
where we have introduced as a shorthand for the quantum numbers 
$\alpha^{\prime}=(s^{\prime}, m^{\prime},t^{\prime},m_{\gamma}^{\prime},m_d^{\prime})$. 
$\rho^{\gamma}_{m_{\gamma}m_{\gamma}^{\prime}}$ and
$\rho^{d}_{m_dm_{d}^{\prime}}$ denote the density matrices of initial
photon polarization and deuteron orientation, respectively, 
$\vec{\Omega}_{s^{\prime}m^{\prime}sm}$ is an operator associated with
the observable, which acts in the two-nucleon spin space and $\rho_s$ 
is a phase space factor given in (\ref{rhos}). For further details we refer 
to~\cite{Aren88,Rob74}.

As is shown in~\cite{Aren88} all polarization observables can be expressed
in terms of the quantities
\beqa
V_{IM} & = &
\frac{1}{\sqrt{3}}~\sum_{m_d^{\prime}m_d}~~\sum_{smt,m_{\gamma}}(-)^{1-m_d^{\prime}}  
~\sqrt{2I+1}~ \left( \begin{array}{ccc}  1 & 1 & I \\
m_d & -m_d^{\prime} & -M \end{array} \right) \nonumber \\ 
 & & \times \int d\Omega_{p_{NN}}~ \rho_s ~ \mathcal M^{(t\mu)~\star}_{sm,m_{\gamma}m_d}~~\mathcal M^{(t\mu)}_{sm,m_{\gamma}m_d^{\prime}}\,,
\label{VIM}
\eeqa
and
\beqa
W_{IM} & = &
\frac{1}{\sqrt{3}}~\sum_{m_d^{\prime}m_d}~~\sum_{smt,m_{\gamma}}(-)^{1-m_d^{\prime}} 
~\sqrt{2I+1}~ \left( \begin{array}{ccc}  1 & 1 & I \\
m_d & -m_d^{\prime} & -M \end{array} \right) \nonumber \\
 & & \times \int d\Omega_{p_{NN}}~\rho_s ~\mathcal M^{(t\mu)~\star}_{sm,m_{\gamma}m_d}~~\mathcal M^{(t\mu)}_{s-m,m_{\gamma}-m_d^{\prime}}\,,
\eeqa
with a Wigner $3j$ symbol. These quantities have the symmetry properties
\beqa
V_{IM}^{\star} & = & (-1)^M ~V_{I-M}\,, \nonumber \\
W_{IM}^{\star} & = & (-1)^I ~W_{IM}\,.
\eeqa

The unpolarized differential cross section is then 
given by
\beqa
\frac{d^3\sigma}{d\Omega_{\pi}dq} & = & V_{00}\,.
\eeqa
The photon asymmetry for linearly polarized photons is given by
\beqa
\Sigma ~\frac{d^3\sigma}{d\Omega_{\pi}dq} & = & - W_{00}\,.
\eeqa
The vector target asymmetry is given by
\beqa
T_{11}~\frac{d^3\sigma}{d\Omega_{\pi}dq} & = & 2~\Im m V_{11}\,.
\label{T11}
\eeqa
The tensor target asymmetries are given by
\beqa
T_{2M} ~\frac{d^3\sigma}{d\Omega_{\pi}dq} & = & (2-\delta_{M0})~\Re e V_{2M}\,,~~
  (M=0,1,2)\,.
\eeqa
The photon and target double polarization asymmetries are given by 

(i) Circular
\beqa
T_{1M}^c ~\frac{d^3\sigma}{d\Omega_{\pi}dq} & = & (2-\delta_{M0})~\Re e
  V_{1M}\,,~~(M=0,1)\,,\nonumber
\eeqa
\vspace*{-1cm}
\beqa
T_{2M}^c ~\frac{d^3\sigma}{d\Omega_{\pi}dq} & = & 2~\Im m V_{2M}\,,~~(M=0,1,2)\,,
\eeqa
(ii) Longitudinal
\beqa
T_{1M}^{\ell} ~\frac{d^3\sigma}{d\Omega_{\pi}dq} & = & i ~W_{1M}\,,~~(M=0,\pm 1)\,,\nonumber
\eeqa
\vspace*{-1cm}
\beqa
T_{2M}^{\ell}~\frac{d^3\sigma}{d\Omega_{\pi}dq} & = & - W_{2M}\,,~~(M=0,\pm 1,\pm 2)\,.
\eeqa
Explicit expressions for unpolarized differential cross section and single 
polarization observables which are predicted and 
discussed in this work are given in terms of the transition matrix elements in 
Appendix~\ref{appen1}.

\section{Results and discussions}
\label{sec6}
In order to do calculations for pion photoproduction on the deuteron we have 
to chosen two ingredients for our model: the deuteron wave function and the 
operator for pion production on a single nucleon.

A large variety of deuteron wave functions is available. They range from 
simple Hulth\'en or Yamaguchi-type wave functions to wave functions obtained 
from modern $NN$ potentials. The contribution to the pion 
production amplitude in~(\ref{tmat_IA_lab}) 
is evaluated by taking a realistic $NN$ potential model for the deuteron 
wave function. For our calculations we have used the wave function of the 
Paris potential~\cite{La+81}, which is in excellent agreement with 
$NN$ scattering data~\cite{Dar03th}. 

The most important ingredient of the model is the operator for pion 
photoproduction on a single nucleon. This operator is obtained in this work 
by computing the 
nonrelativistic reduction of the amplitudes for the Feynman diagrams in 
Fig.~\ref{Fdiagrams}. As already seen in 
section~\ref{sec2}, that our calculations 
for the elementary process are in good agreement with recent experimental 
data as well as with other theoretical predictions and gave a clear indication 
that this elementary operator is quite satisfactory for our purpose.

The discussion of our results is divided into four parts. First,
we will discuss the $\pi$-meson spectra as a function of the absolute value 
of pion momentum $q$ at forward and backward emission pion angles 
$\theta_{\pi}$ for photon energy at the $\Delta$(1232)-resonance region, 
i.e. $\omega_{\gamma}^{\rm lab}=330$ MeV. In the second part, we will then 
consider the photon asymmetry $\Sigma$ for linearly polarized photons. 
Our results for the vector target asymmetry $T_{11}$ will be presented in 
the third part. In the last part, we will discuss our results for the tensor 
target asymmetries $T_{20}$, $T_{21}$, and $T_{22}$. In all parts, we will give 
a calculations for all the three isospin channels of the  $d(\gamma,\pi)NN$ 
reaction. For comparison, we will always present the results for the full calculation and 
the results when only the $\Delta$(1232)-resonance contribution is taken into account.

All the above mentioned observables are calculated, as seen in 
Appendix~\ref{appen1}, by integrating over the polar 
angle $\theta_{p_{NN}}$ and the azimuthal angle $\phi_{p_{NN}}$ of the 
relative momentum $\vec p_{NN}$ of the two outgoing nucleons. These 
integrations are carried out numerically. The number of integration points 
was being increased until the accuracy of calculated observable becomes good 
to 1$\%$.

\subsection{The $\pi$-meson spectra}
\label{sec61}
\begin{figure}[htb]
\begin{center}
\includegraphics[scale=.7]{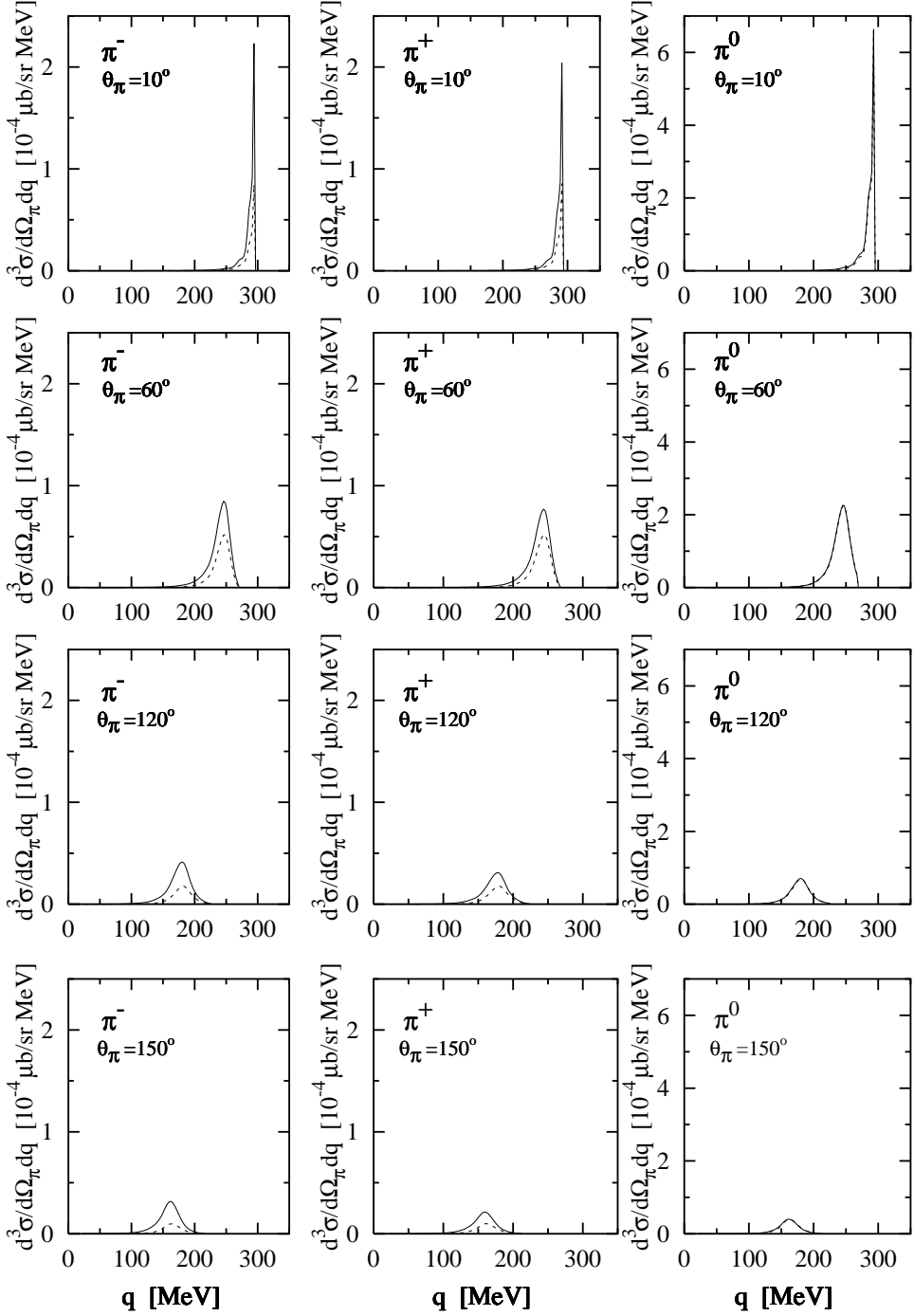}
\caption{The $\pi$-meson spectra in the $d(\gamma,\pi)NN$ reaction as a 
function of the absolute value of pion momentum $q$ at a photon energy of 330 
MeV for four different values of emission pion angles $\theta_{\pi}$. The solid curves 
show the results of the full calculations while the dotted curves represent 
the results when only the $\Delta$(1232)-resonance is taken into account. 
The left, middle and right panels represent the results for 
$\gamma d\to\pi^-pp$, $\pi^+nn$ and $\pi^0np$, respectively.}  
\label{unpolcs1}
\end{center}
\end{figure}

We start our discussion with the $\pi$-meson spectra, i.e., the unpolarized 
differential cross section $d^3\sigma/(d\Omega_{\pi}dq)$ which comes from the 
fully exclusive differential cross section 
$d^{5}\sigma/(d\Omega_{\pi}dqd \Omega_{p_{NN}})$ by integrating over 
$\Omega_{p_{NN}}$. In Fig.~\ref{unpolcs1} we depict our results for the $\pi$-meson spectra as a 
function of the absolute value of pion momentum $q$ at four different values 
of emission pion angles $\theta_{\pi}$ for each isospin channel of the 
$\gamma d\to\pi NN$ reaction for $\omega_{\gamma}^{lab}=330$ MeV. One sees, 
that when the absolute value of pion momentum $q$ reaches 
its maximum, the absolute value of the relative momentum $p_{NN}$ 
of the two outgoing nucleons vanishes, and thus a narrow 
peak is appears in the forward emission pion angles for charged as well as for neutral 
pion photoproduction channels. In the lower part of Fig.~\ref{unpolcs1} we see, that the unpolarized differential cross section is small 
and the narrow peak which appears at forward emission pion angles is disappears. The 
same effect appears in the coherent process of charged pion photo- and 
electroproduction on the deuteron~\cite{Lag78,Kob87}, in deuteron 
electrodisintegration~\cite{Fab76} as well as in 
$\eta$-photoproduction~\cite{Fix97}. It is also clear that the maximum value 
of $q$ (when $p_{NN}\to 0$) is decreases with increasing the emission pion 
angle. It changes from $\sim$ 300 MeV at forward 
emission angles to $\sim$ 200 MeV at backward ones. In principle, the experimental observation of this peak in the high 
$\pi$-momentum spectrum may serve as another evidence for the understanding 
of the $\pi$-meson spectra.

In conclusion, one notes that the contributions from Born terms are important 
for charged pion production channels but these are much less important in the 
case of neutral pion production.

\subsection{Photon asymmetry}
\label{sec62}
\begin{figure}[htb]
\begin{center}
\includegraphics[scale=.7]{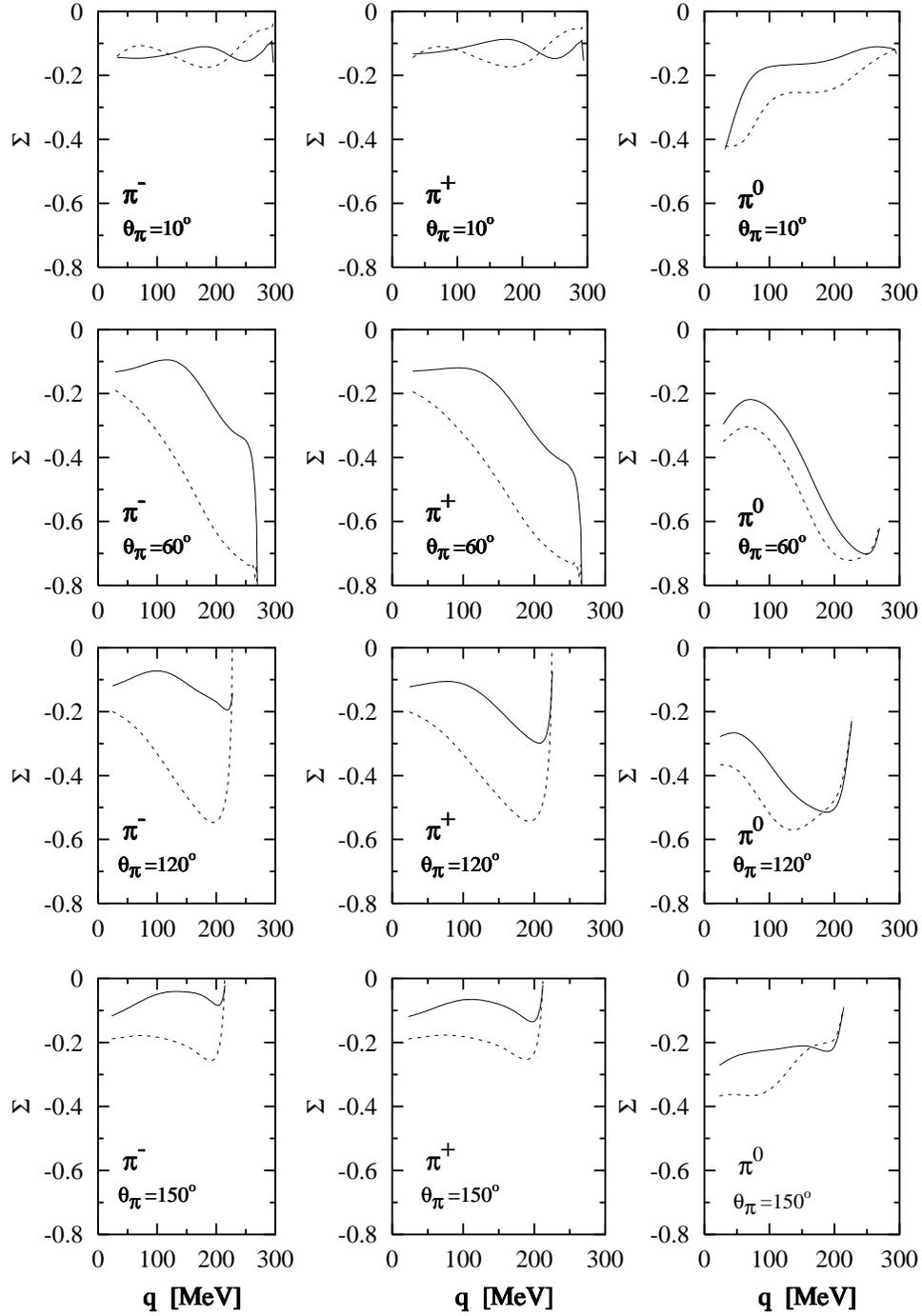}
\caption{Photon asymmetry $\Sigma$ of 
  $d(\vec\gamma,\pi)NN$. Notation of the curves as in Fig.~\ref{unpolcs1}.}
\label{phasym1}
\end{center}
\end{figure}

Here we discuss our results for the photon asymmetry $\Sigma$ for 
linearly polarized photons for all the different charge 
states of the pion of $d(\vec\gamma,\pi)NN$. The $\gamma$-asymmetry for 
fixed pion angles of 
$10^{\circ}$, $60^{\circ}$, $120^{\circ}$, and $150^{\circ}$ are plotted in 
Fig.~\ref{phasym1} as a function of the absolute value of pion momentum 
$q$ at $\omega_{\gamma}^{lab}=330$ MeV. The dotted curves show 
the contribution of the $\Delta$(1232)-resonance alone in order to clarify the 
importance of the Born terms. First of all, we see that the photon asymmetry 
has always a negative values at forward and backward emission pion angles for 
charged as well as for neutral pion channels. One notes qualitatively a 
similar behaviour for charged pion channels whereas a totally different 
behaviour is seen for the neutral pion channel.

For extreme forward and backward pion angles one sees, that the 
effect of Born contributions is relatively small in comparison to the 
results at $\theta_{\pi}=60^{\circ}$ and $120^{\circ}$. At $60^{\circ}$ we see 
a strong reduction of the photon asymmetry at maximum pion momentum. This 
reduction changes to an overestimation at backward pion angles. One notices also, 
that the contributions from Born terms are much important, in particular at 
$q\simeq 200$ MeV which is very clear for charged pion channels. 
We observe that the interference of the Born terms with the 
$\Delta$(1232)-resonance contribution causes considerable changes in the 
photon asymmetry. Experimental measurements as well as other theoretical 
predictions will give us more valuable information on the photon asymmetry.

\subsection{Vector target asymmetry}
\label{sec63}
\begin{figure}[htb]
\begin{center}
\includegraphics[scale=.7]{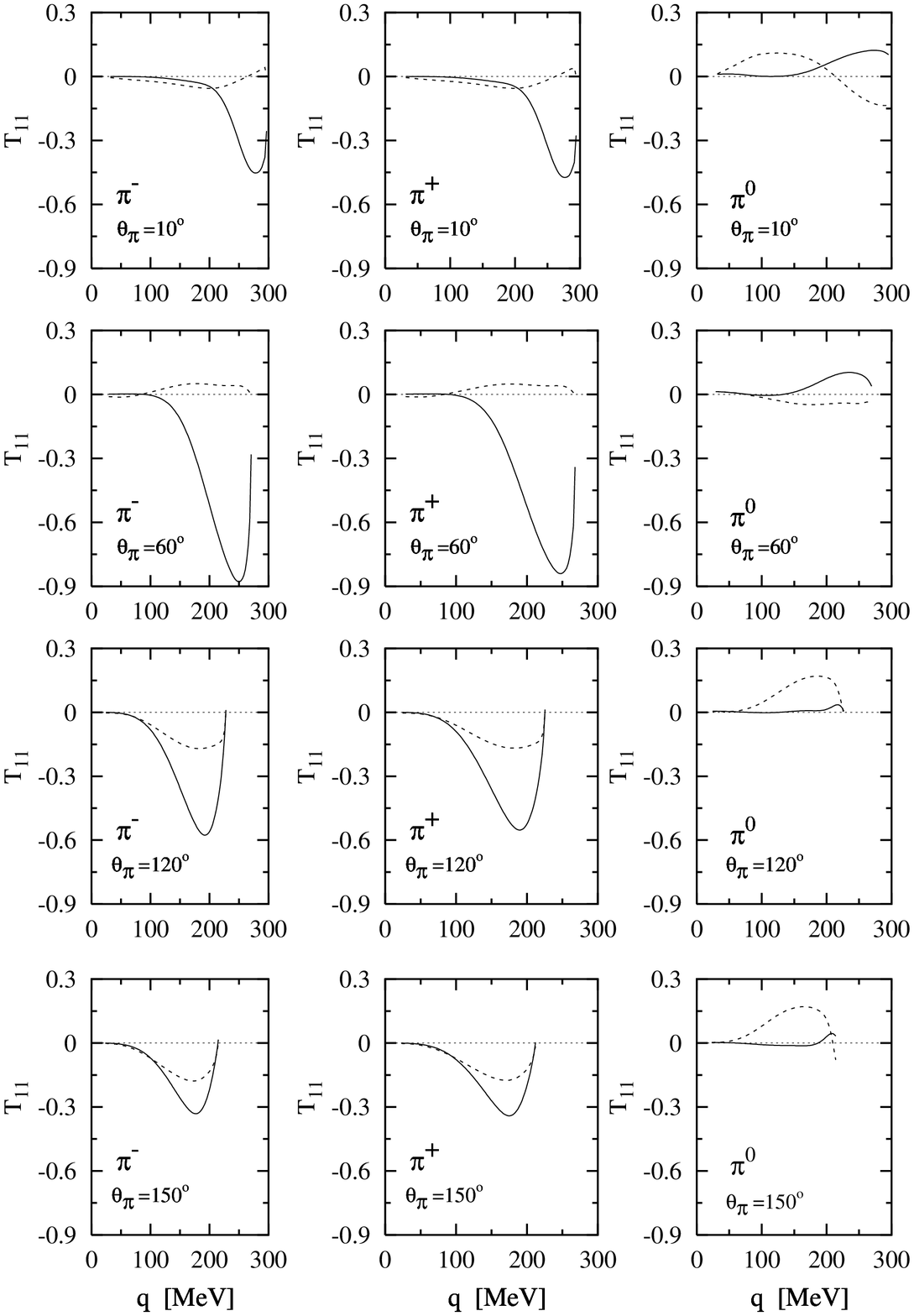}
\caption{Vector target asymmetry $T_{11}$ of $\vec d(\gamma,\pi)NN$. 
  Notation of the curves as in Fig.~\ref{unpolcs1}.}
\label{vtasym1}
\end{center}
\end{figure}

In this subsection we present and discuss our results for the vector target 
asymmetry $T_{11}$. Fig.~\ref{vtasym1} shows these results as a function of 
the absolute value of pion momentum $q$ at four different values of pion angles 
$\theta_{\pi}$ for $\omega_{\gamma}^{lab}=330$ MeV. The asymmetry $T_{11}$ 
clearly differs in size between charged and neutral pion photoproduction 
channels, being even opposite in phase. For charged pion 
photoproduction reactions we see from the left and middle panels of 
Fig.~\ref{vtasym1}, that the vector target asymmetry has always a negative 
values. At 
forward pion angles these values come mainly from the Born terms since a small 
contribution from the $\Delta$-resonance was found. At backward 
angles, the negative values come from an interference of the Born terms with 
the $\Delta$(1232)-resonance contribution since the $\Delta$-contribution is 
large in this case. 

With respect to the neutral pion photoproduction reaction, we see from the 
solid curves of the right panel of Fig.~\ref{vtasym1}, that the vector target 
asymmetry $T_{11}$ has a very small negative values at smaller pion momentum 
and a relatively large positive values at higher pion momentum. It is 
interesting to point out the importance of the Born terms in the charged pion 
production reactions in comparison to the contribution of the 
$\Delta$(1232)-resonance. The sensitivity of $T_{11}$ to the Born terms has 
also been discussed by Blaazer {\it et al.}~\cite{Bla94} and Wilhelm and 
Arenh\"ovel~\cite{Wil95} for the coherent pion photoproduction reaction 
on the deuteron. The reason is that $T_{11}$ depends on the relative phase 
of the matrix elements as can be seen from (\ref{VIM}) and (\ref{T11}). 
It would vanish for a constant overall phase of the $t$-matrix, a case which 
is approximately realized if only the $\Delta$(1232)-amplitude is considered.

\subsection{Tensor target asymmetries}
\label{sec64}
\begin{figure}[htb]
\begin{center}
\includegraphics[scale=.7]{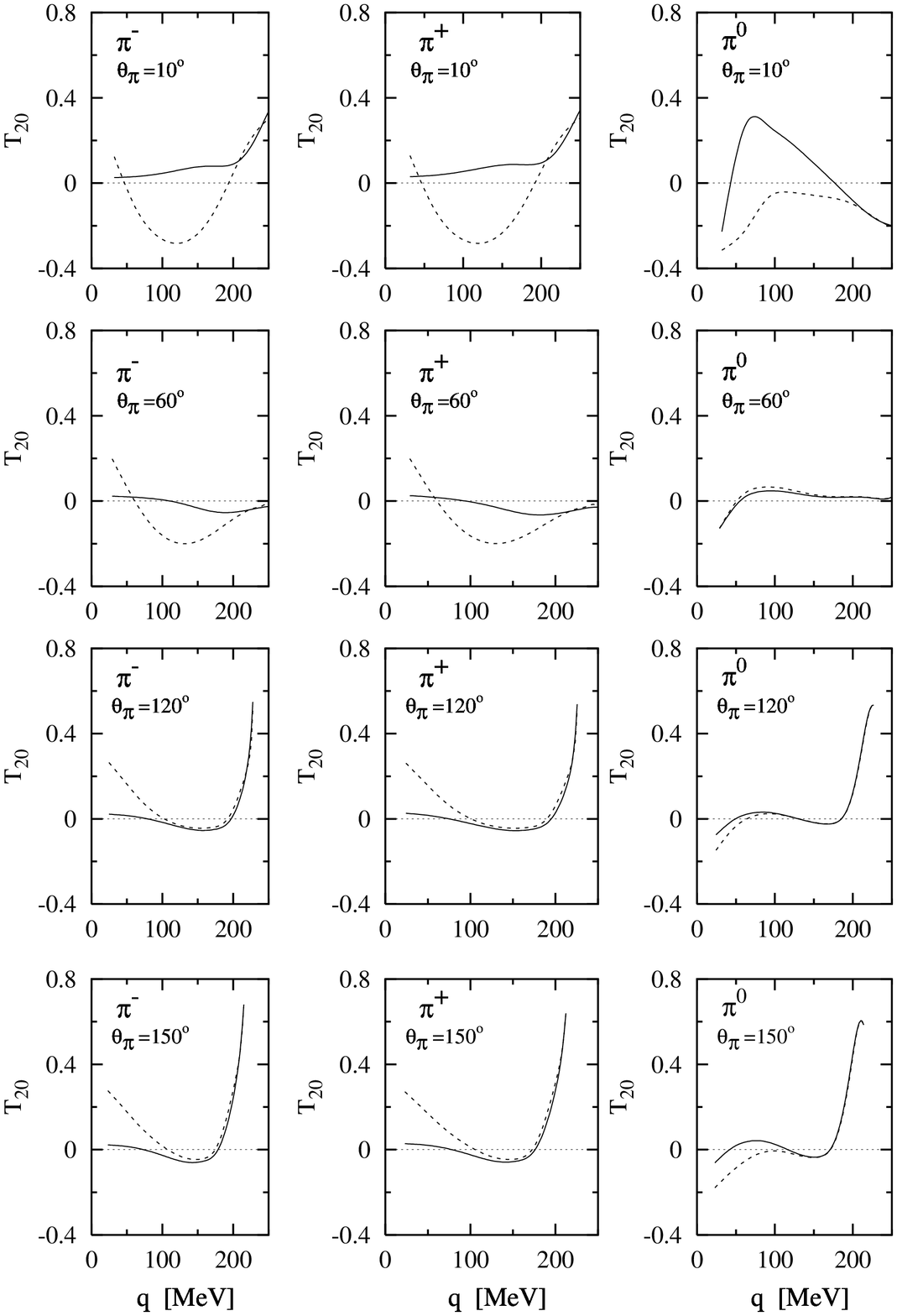}
\caption{Tensor target asymmetry $T_{20}$ of $\vec d(\gamma,\pi)NN$. 
  Notation of the curves as in Fig.~\ref{unpolcs1}.}
\label{ttasym201}
\end{center}
\end{figure}

Let us present and discuss now the results of the tensor target asymmetries 
$T_{20}$, $T_{21}$, and $T_{22}$ for $\vec d(\gamma,\pi)NN$. 
We start from the tensor asymmetry $T_{20}$. For 
$\gamma d\to\pi NN$ at forward and backward emission pion angles, the asymmetry $T_{20}$ 
allows one to draw specific conclusions about details of the reaction 
mechanism. Results for $T_{20}$ are plotted in 
Fig.~\ref{ttasym201} at four different values 
of pion angles as a function of the absolute value of pion momentum $q$ 
for $\omega_{\gamma}^{lab}=330$ MeV. The 
dotted curves represent our results for the contribution of the 
$\Delta$(1232)-resonance and the solid ones show the results when the 
Born terms are included. In general, one notes again the importance 
of Born terms in the case of charged pion photoproduction channels (see the 
left and middle panels of Fig.~\ref{ttasym201}). In the case of neutral pion 
production channel (right panel of Fig.~\ref{ttasym201}) one sees, that the 
Born terms are important only at extreme forward pion angles. One sees 
also that the contribution of Born terms is very small for backward pion 
angles and higher pion momentum, but it is relatively large for small pion 
momentum.

\begin{figure}[htb]
\begin{center}
\includegraphics[scale=.7]{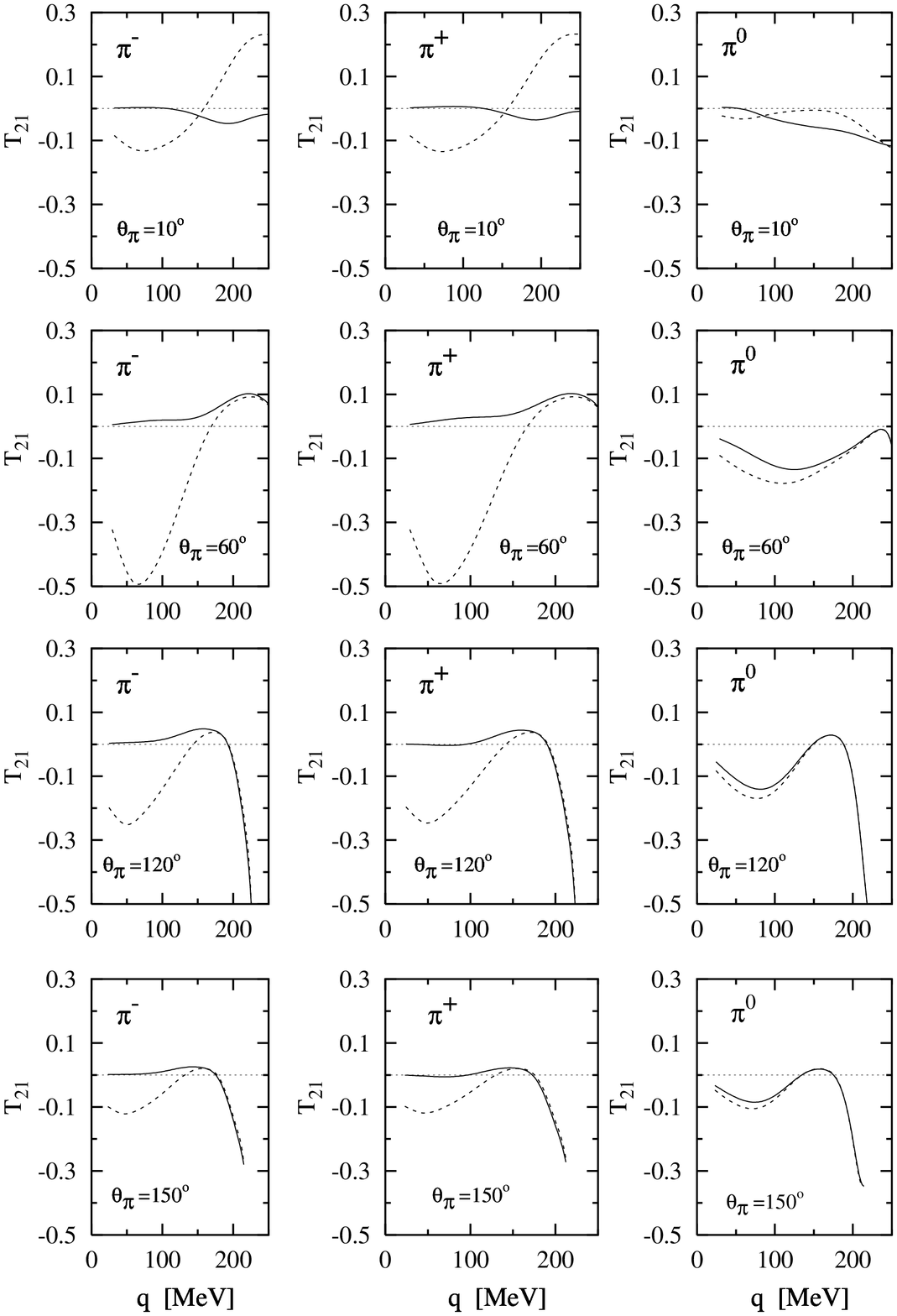}
\caption{Tensor target asymmetry $T_{21}$ of $\vec d(\gamma,\pi)NN$. 
  Notation of the curves as in Fig.~\ref{unpolcs1}.}
\label{ttasym211}
\end{center}
\end{figure}

Fig.~\ref{ttasym211} shows our results for the tensor target asymmetry 
$T_{21}$ as a function of $q$ for fixed pion angles 
$\theta_{\pi}=10^{\circ}$, $60^{\circ}$, $120^{\circ}$, and $150^{\circ}$ at 
$\omega_{\gamma}^{\rm lab}=330$ MeV. The dotted curves are 
due to calculations done without the inclusion of the Born terms. One notices 
that the $T_{21}$ asymmetry is sensitive to Born terms, in particular at 
forward pion angles. Also in this case one notes the importance of Born terms 
in the case of charged pion photoproduction reactions, in particular at 
smaller pion momentum. In the case of $\pi^0$ channel one sees that the 
contribution of Born terms is much less important at all angles.

\begin{figure}[htb]
\begin{center}
\includegraphics[scale=.7]{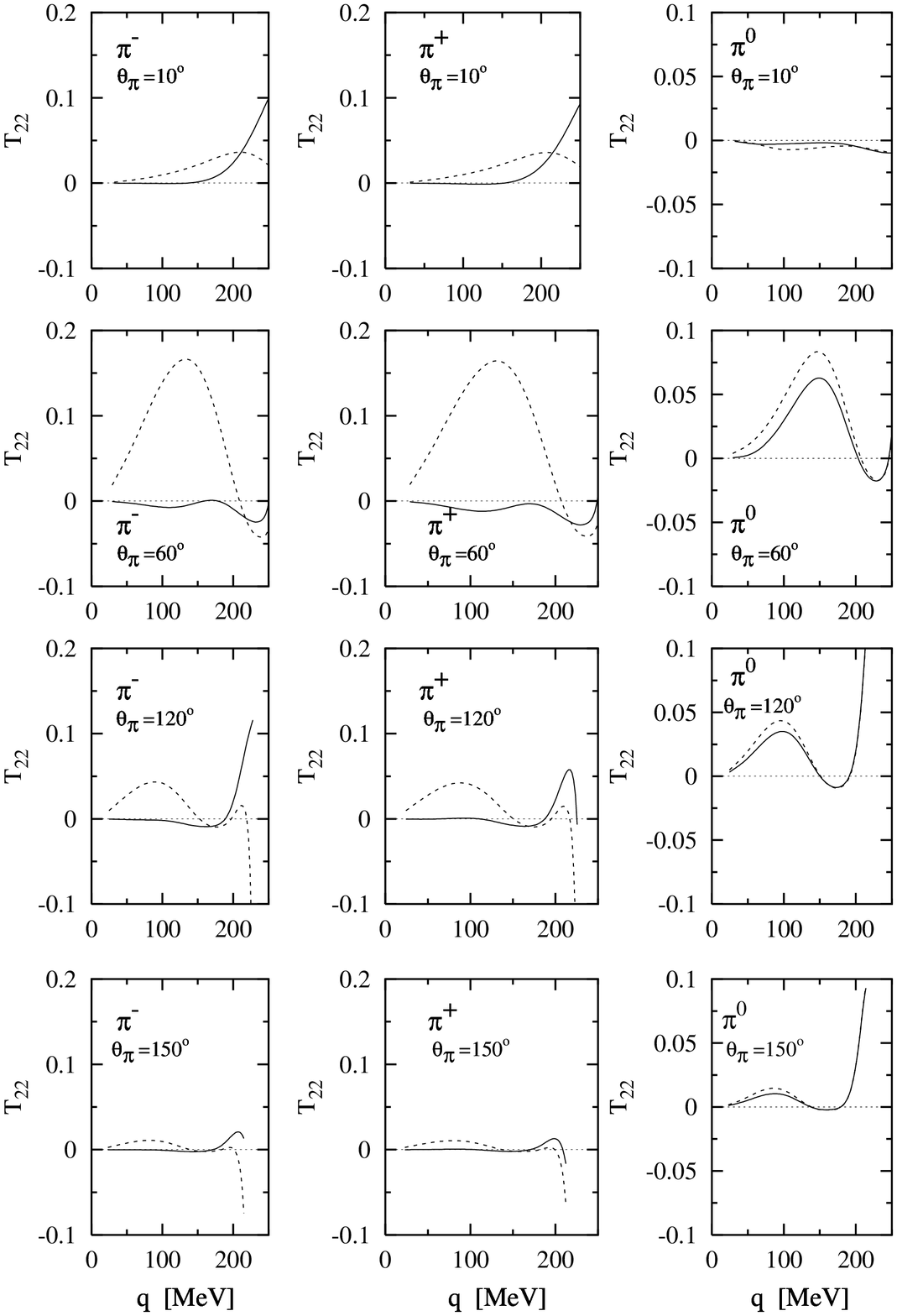}
\caption{Tensor target asymmetry $T_{22}$ of $\vec d(\gamma,\pi)NN$. 
  Notation of the curves as in Fig.~\ref{unpolcs1}.}
\label{ttasym221}
\end{center}
\end{figure}

In Fig.~\ref{ttasym221} we depict our results for the tensor target asymmetry 
$T_{22}$ as a function of $q$. We have used here the same 
four values of pion angle $\theta_{\pi}$ as in the previous figures. One readily 
notes the importance of Born terms, in particular for charged pion channels. Like 
the results of Figs.~\ref{ttasym201} and \ref{ttasym211}, the $T_{22}$ asymmetry is sensitive to the values of 
pion angle $\theta_{\pi}$. We notice that the $T_{22}$ asymmetry changes 
dramatically if only the $\Delta$-contribution is taken into account. 
At $\theta_{\pi}=60^{\circ}$ we see that the Born terms are 
very important, especially for charged pion photoproduction channels.
In the case of neutral pion production channel these terms are much less 
important.

\section{Conclusion}
\label{sec7}
In this paper we have studied incoherent single pion photoproduction 
on the deuteron in the $\Delta$(1232)-resonance region with special 
emphasis on polarization observables. The $\gamma d\to\pi NN$ scattering 
amplitude is given as a linear combination of the on-shell matrix elements 
of pion photoproduction on the two nucleons. For the elementary pion 
photoproduction operator an effective Lagrangian model is used which is 
based on time-ordered perturbation theory and describes well the elementary
$\gamma N\to\pi N$ reaction.

Particular attention was paid to $\pi$-meson spectra as well as single 
polarization observables. We have 
presented results for the unpolarized differential cross section 
$d^3\sigma/d\Omega_{\pi}dq$, photon asymmetry $\Sigma$ for linearly 
polarized photons, vector target asymmetry $T_{11}$ and tensor target 
asymmetries $T_{20}$, $T_{21}$, and $T_{22}$. As already noticed in the discussion above, 
we found that interference of 
Born terms and the $\Delta$(1232)-contribution plays a significant role. 
Unfortunately, there are no experimental data available to be compared to the 
observables we computed.

We would like to conclude that the results presented here for polarization 
observables in the $\d(\gamma,\pi)NN$ reaction in the $\Delta$-resonance 
region can be used as a basis for the simulation of the behaviour of 
polarization observables and for an optimal planning of new polarization 
experiments of this reaction. It would be very interesting to examine our 
predictions experimentally.

Finally, we would like to point out that future improvements of the present 
model should include further investigations including a three-body treatment 
of the final $\pi NN$ system for the lowest and most important partial waves. 
As future refinements we consider also the use of a more sophisticated 
elementary production operator, which will allow one to extend the present 
approach to higher energies, and the role of irreducible two-body 
contributions to the electromagnetic pion production operator. 

\begin{appendix}
\section{Explicit expressions for polarization observables}
\label{appen1}
In this appendix we give the formal expressions for unpolarized differential 
cross section and single polarization 
observables which are presented and discussed in this paper. 
The 3-fold unpolarized differential cross section is obtained from the fully 
exclusive differential cross section (\ref{fivefold}) by integration over 
$\Omega_{p_{NN}}$
\beqa
\label{unpdcs}
\frac{d^3\sigma}{d\Omega_{\pi} dq} & = & \frac 16 ~\mathcal F
\eeqa
with
\beqa
\label{fff}
\mathcal F &=& \sum_{\alpha} \int d\Omega_{p_{NN}} 
  ~\rho_{s}~ |{\mathcal M}^{(t\mu)}_{sm m_{\gamma}m_d}|^{2}\,.
\eeqa

The photon asymmetry for linearly polarized photons is given by
\beqa
\Sigma ~& = &~ \frac{d\sigma^{\parallel} - d\sigma^{\perp}}
{d\sigma^{\parallel} + d\sigma^{\perp}}\nonumber \\
 & = & \frac{2}{\mathcal F} ~\Re e \sum_{s,m,t,m_{d}} 
\int d\Omega_{p_{NN}} ~\rho_{s} ~{\mathcal M}^{(t\mu)}_{sm +1m_d} 
~{\mathcal M}^{(t\mu)~\star}_{sm -1m_d}\,,
\label{FSigma}
\eeqa
where $d\sigma^{{\parallel}({\perp})}$ is the differential cross section 
for incoming photons polarized parallel (perpendicular) to the reaction 
plane.

The vector target asymmetry $T_{11}$ is given by
\beqa
T_{11} & = & \frac{\sqrt{6}}{\mathcal F} ~\Im m\sum_{s,m,t,m_{\gamma}} 
\int d\Omega_{p_{NN}}~ \rho_{s}\Big[{\mathcal M}^{(t\mu)}_{smm_{\gamma}-1} 
- {\mathcal M}^{(t\mu)}_{smm_{\gamma}+1}\Big]
{\mathcal M}^{(t\mu)~\star}_{smm_{\gamma}0}\,.
\label{FT11}
\eeqa

The tensor target asymmetries are expressed in terms of the amplitudes as follows
\beqa
T_{20} & = & \frac{1}{\sqrt{2}\mathcal F} \sum_{s,m,t,m_{\gamma}} 
\int d\Omega_{p_{NN}}~\rho_{s}~\Big[~|{\mathcal M}^{(t\mu)}_{smm_{\gamma}+1}|^2 
+ |{\mathcal M}^{(t\mu)}_{smm_{\gamma}-1}|^2\nonumber \\
& & \hspace{4.8cm}- 2~ |{\mathcal M}^{(t\mu)}_{smm_{\gamma}0}|^2~\Big]\,,
\label{FT20}
\eeqa
\beqa
T_{21} & = & \frac{\sqrt{6}}{\mathcal F} ~\Re e\sum_{s,m,t,m_{\gamma}} 
\int d\Omega_{p_{NN}} ~\rho_{s}~ \Big[{\mathcal M}^{(t\mu)}_{smm_{\gamma}-1} 
- {\mathcal M}^{(t\mu)}_{smm_{\gamma}+1}\Big] ~
{\mathcal M}^{(t\mu)~\star}_{smm_{\gamma}0}\,, 
\label{FT21}
\eeqa
\beqa
T_{22} & = & \frac{2\sqrt{3}}{\mathcal F} ~\Re e\sum_{s,m,t,m_{\gamma}} 
\int d\Omega_{p_{NN}} ~\rho_{s}~ {\mathcal M}^{(t\mu)}_{smm_{\gamma}-1} 
~{\mathcal M}^{(t\mu)~\star}_{smm_{\gamma}+1} \,. 
\label{FT22}
\eeqa
\end{appendix}

\begin{ack}
E.M.\ Darwish is grateful to Professor H.\ Arenh\"ovel as well as many 
scientists of the Institut f\"ur Kernphysik of the Johannes 
Gutenberg-Universit\"at, Mainz for the very kind hospitality and fruitful 
discussions.
\end{ack}


\newpage

\vspace*{1cm}

\end{document}